\documentclass[english,pra,aps, superscriptaddress,twocolumn,longbibliography]{revtex4-1}
\usepackage[T1]{fontenc}
\usepackage[latin9]{inputenc}
\usepackage{mathrsfs}
\usepackage{booktabs}
\usepackage{amsmath}
\usepackage{amssymb}
\usepackage{graphicx}
\usepackage{esint}

\makeatletter

\usepackage[colorlinks,citecolor=blue,linkcolor=blue]{hyperref}
\providecommand{\tabularnewline}{\\}

\makeatother

\usepackage{babel}
\begin{document}
\title{Effect of finite-size heat source's heat capacity on the efficiency
of heat engine}
\author{Yu-Han Ma}
\email{yhma@csrc.ac.cn}

\address{Beijing Computational Science Research Center, Beijing 100193, China}
\address{Graduate School of China Academy of Engineering Physics, No. 10 Xibeiwang
East Road, Haidian District, Beijing, 100193, China}
\begin{abstract}
Heat engines used to output useful work have important practical significance,
which, in general, operate between heat baths of infinite size and
constant temperature. In this paper we study the efficiency of a heat
engine operating between two finite-size heat sources with initial
temperature differences. The total output work of such heat engine
is limited due to the finite heat capacity of the sources. We investigate
the effects of different heat capacity characteristics of the sources
on the heat engine's efficiency at maximum work (EMW) in the quasi-static
limit. In addition, we study the efficiency of the engine working
in finite-time with maximum power of each cycle is achieved and find
the efficiency follows a simple universality as $\eta=\eta_{\mathrm{C}}/4+O\left(\eta_{\mathrm{C}}^{2}\right)$.
Remarkably, when the heat capacity of the heat source is negative,
such as the black holes, we show that the heat engine efficiency during
the operation can surpass the Carnot efficiency determined by the
initial temperature of the heat sources. It is further argued that
the heat engine between two black holes with vanishing initial temperature
difference can be driven by the energy fluctuation. The corresponding
EMW is proved to be $\eta_{\mathrm{EMW}}=2-\sqrt{2}$, which is two
time of the maximum energy release rate $\mu=\left(2-\sqrt{2}\right)/2\approx0.29$
of two black hole emerging process obtained by S. W. Hawking.
\end{abstract}
\maketitle

\section{introduction}

As one of the most useful devices in modern society, the heat engine
converts the heat extracted from the heat source into useful work,
which is one of the core fields in thermodynamic research \citep{Carnoteff,Esposito_2009,Campisi_2011,kosloff2014quantum,Pekola_2015,Vinjanampathy_2016,Binder2018,kosloff2019quantum}.
Early heat engine research was limited to reversible cycles in the
quasi-static limit , with which, as stated by the Carnot's theorem
\citep{Carnoteff}, the achievable maximum efficiency of heat engines
is the so-called Carnot efficiency $\eta_{\mathrm{C}}=1-T_{\mathrm{L}}/T_{\mathrm{H}},$where
$T_{\mathrm{H}}$ ($T_{\mathrm{L}}$) is the temperature of the hot
(cold) bath. Since the last century, with the maturity of quantum
theory and its related technologies, people started to pay attention
to the performance of quantum heat engines working in micro-scale
within the framework of quantum thermodynamics\citep{Binder2018,kosloff2019quantum,kosloff2014quantum,Campisi_2011,scovil1959three,alicki1979quantum,kosloff1984quantum,scully2003extracting,QTquanQH,Brantut2013,Dechant2015,Rossnagel2016,shortcut2adiabaticityLutz,passos2019optical}.
Series of the quantum effect, such as coherence, entanglement, quantum
phase transition, etc., of the working substance or heat source have
been studied to realize better heat engines \citep{scully2003extracting,IsolatedquantumHE,NanoscaleHEPRL2014,YHMaQPTHE,Brandner2017,Su2018,dorfman2018efficiency,Camati2019,chen2019boosting}.
On the other hand, with the development of non-equilibrium thermodynamics\citep{de2013non,Esposito_2009},
the optimization of actual heat engines under the framework of finite-time
thermodynamics attracted a wide range of attention \citep{andresen1983finite,andresen1984thermodynamics,wu1999recent,ZCTuCPB,Holubec2017}.Extensive
research on the efficiency at maximum power (EMP) \citep{CA,Andresen1977,chen1994maximum,SekimotoJPSJ,BroeckPRL2005,EspositoPRL2010,Tu2008JPhysAMathTheor41_312003},
trade-off relation between power and efficiency \citep{tradeoffholubec,TradeoffrelationShiraishi,CavinaPRLtradeoffrelation,Constraintrelationyhma},
and optimal operation of heat engine have been proposed \citep{yhmaoptimalcontrol,Ma2019IEGExp}.
The motivation for these studies stems from the fact that time is
a finite resource, we cannot trade at the cost of infinitely long
working time for heat engines that are efficient but have vanishing
output power.
\begin{widetext}
\begin{center}
\begin{table}
\centering{}%
\begin{tabular}{ccc}
\toprule 
\addlinespace[0.2cm]
 & $t\rightarrow\infty$ & $t\nrightarrow\infty$\tabularnewline\addlinespace[0.2cm]
\midrule 
\addlinespace[0.2cm]
$C\rightarrow\infty$ & $\eta_{\mathrm{max}}=\eta_{\mathrm{C}}$ & $\frac{\eta_{\mathrm{C}}}{2}\leq\eta_{\mathrm{EMP}}\leq\frac{\eta_{\mathrm{C}}}{2-\eta_{\mathrm{C}}}$\tabularnewline\addlinespace[0.2cm]
\midrule 
\addlinespace[0.2cm]
$C\nrightarrow\infty$ & $1+\frac{\left(1-\eta_{\mathrm{C}}\right)\ln\left(1-\eta_{\mathrm{C}}\right)}{\eta_{\mathrm{C}}}\leq\eta_{\mathrm{EMW}}\leq1+\frac{\eta_{\mathrm{C}}}{\ln\left(1-\eta_{\mathrm{C}}\right)}$ & $1+\frac{\eta_{\mathrm{C}}/2}{\ln\left(1-\eta_{\mathrm{C}}/2\right)}\leq\eta^{\mathrm{FT}}\leq1+\frac{1-\eta_{\mathrm{C}}}{\eta_{\mathrm{C}}/2}\ln\frac{1-\eta_{\mathrm{C}}}{1-\eta_{\mathrm{C}}/2}$\tabularnewline\addlinespace[0.2cm]
\bottomrule
\end{tabular}\caption{\label{tab:Bound-for-efficiency}Bound for efficiency in different
case. Here $t$ is the operation time of the heat engine and $C$
is the heat capacity of the heat source. In the case engine working
in quasi-static cycle between infinite heat bath, i.e., $t\rightarrow\infty$,$C\rightarrow\infty$,
the maximum achievable efficiency, as stated by Carnot, is the Carnot
efficiency $\eta_{\mathrm{C}}=1-T_{\mathrm{L}}/T_{\mathrm{H}}$. For
the engine operates in finite time, i.e., $C\rightarrow\infty$, $t\nrightarrow\infty$,
Esposito et al. \citep{EspositoPRL2010} give the bounds for the efficiency
at maximum power (EMP) with low-dissipation Carnot-like engine. We
show the bounds for efficiency at maximum work (EMW) ($t\rightarrow\infty$,
$C\nrightarrow\infty$) and efficiency at maximum power for each cycle
($t\nrightarrow\infty$, $C\nrightarrow\infty$) obtained in this
paper. The detailed derivations are illustrated in Sec. \ref{subsec:High-temperature-limit}
and Sec. \ref{sec:Efficiency-at-maximum} respectively. The bounds
in the latter two cases for $\eta_{\mathrm{EMW}}$ and $\eta^{\mathrm{FT}}$
in this table are limited to the heat source having a positive and
constant heat capacity. The bounds correspond to heat capacity change
with temperature are discussed in Sec. \ref{subsec:low-temperature-limit}
while the negative heat capacity case are discussed in Sec. \ref{sec:Heat-source-with-nagetive}.}
\end{table}
\par\end{center}

\end{widetext}

Most of studies about heat engines have regarded the thermal source
as an infinite system, that is, it can continuously provide heat.
However, just like time, the heat source is also a finite resource,
so it is an interesting and practical task to consider the optimization
of the heat engine working between finite-size heat sources. Recently,
people began to consider this issue with different perspectives. For
example, considering the Carnot heat engine working between finite
heat sources \citep{ondrechen1981maximum,ondrechen1983generalized,leff1987available},
linear irreversible heat engines working in finite time with finite-size
bath \citep{izumida2014work,wang2014optimization}, and the bounds
of optimal efficiency the engines can achieve \citep{johal2016optimal,johal2016near}.
In addition, the influences of the finite-size heat source on the
quantum heat engine \citep{tajima2017finite,sparaciari2017resource,richens2018finite,pozas2018quantum,mohammady2019efficiency}
and quantum battery \citep{barra2019dissipative} also attract some
attention. In general, the finite-size effect of the heat source is
reflected in the limited heat capacity. Therefore, the nature of the
heat capacity of the heat sources directly determines the performance
of the heat engine working between them. In this paper, we discuss
the effects of finite-size heat sources with different heat capacity
characteristics on heat engine's efficiency, in both quasi-static
and finite-time circumstances. We obtain some bounds for the efficiency
and list them in Tab. \ref{tab:Bound-for-efficiency}. In particular,
we study the case where the heat source has a negative heat capacity,
which is proved to be advantageous for improving the efficiency of
the heat engine. To our best knowledge, this has never been reported
before.

The paper is organized as follows: In Sec. \ref{sec:heat-engine-working-finite source},
we first generally discuss the influence of the heat capacity of the
finite-size heat sources on the efficiency of the heat engine at maximum
work (EMW) with quasi-static cycle. Then the efficiency of the heat
engine in the high and low temperature limit with different heat capacity
function are derived. In the low-temperature regime, it is found that
the dimension of the heat source will influence the EMW of the heat
engine, and the higher EMW can be achieved with higher dimension materials.
In Sec. \ref{sec:Efficiency-at-maximum}, the study in quasi-static
situation is extend to the finite-time case, where we consider the
low-dissipation Carnot-like heat engine working between two finite-size
sources. We point out that when the heat engines operates with maximum
power in each cycle, the efficiency of the whole process follows a
simple universality as $\eta=\eta_{\mathrm{C}}/4+O\left(\eta_{\mathrm{C}}^{2}\right)$.
With the black hole as an illustration, we study the negative heat
capacity system service as the finite-size heat source in Sec. \ref{sec:Heat-source-with-nagetive}.
Conclusion and discussions are given in Sec. \ref{sec:Conclusion}.

\section{\label{sec:heat-engine-working-finite source}heat engine working
between finite-size heat source}

\begin{figure}
\includegraphics[width=8.5cm]{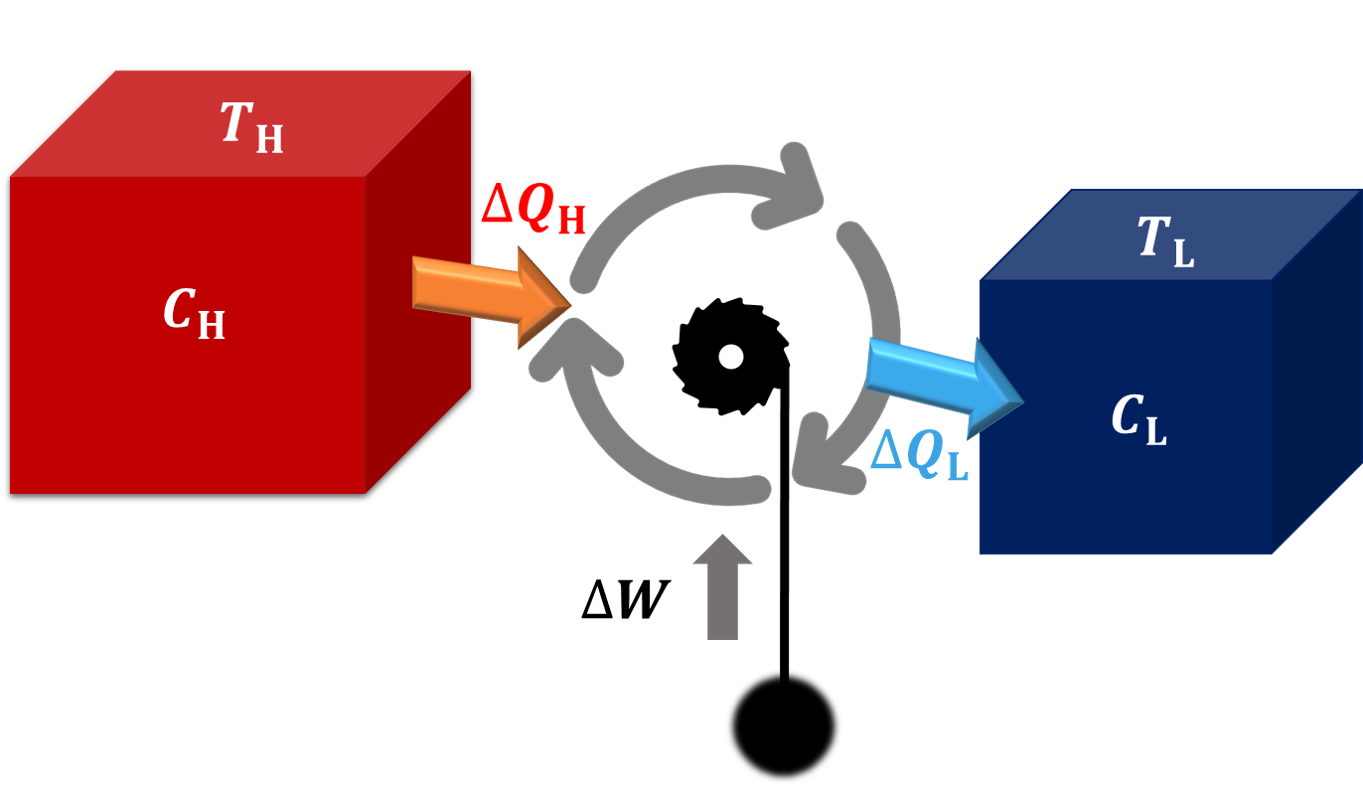}\caption{Heat engine working between two finite-size heat sources. $T_{\mathrm{H}}$
($T_{\mathrm{L}}$) and $C_{\mathrm{H}}$ ($C_{\mathrm{L}}$) are
the initial temperature and heat capacity of the high (low) temperature
source $\mathrm{H}$ ($\mathrm{L}$) respectively. $\Delta W=\sum_{\mathrm{\alpha=H},\mathrm{L}}\Delta Q_{\alpha}$
is the output work of the engine per cycle while $\Delta Q_{\mathrm{H}}$
is the heat absorbed from the hot source and $\Delta Q_{\mathrm{L}}$
the heat releases to the cold source.}

\end{figure}

As shown in Fig. 1, we consider the heat engine of interest is working
between two heat source $\mathrm{H}$ and $\mathrm{L}$ with finite
size, where $T_{\mathrm{H}}$ ($T_{\mathrm{L}}$) and $C_{\mathrm{H}}$
($C_{\mathrm{L}}$) are the initial temperature and heat capacity
of the high (low) temperature source $\mathrm{H}$ ($\mathrm{L}$)
respectively. With the working substance S run through an arbitrary
thermodynamic cycle, the heat engine can generate output work $\Delta W=\int_{0}^{\tau}dW$
per cycle. Here $\tau$ is the cycle time, and $\Delta W=\sum_{\mathrm{\alpha=H},\mathrm{L}}\Delta Q_{\alpha}$,
$\Delta Q_{\mathrm{H}}=-\int_{0}^{\tau}dQ_{\mathrm{H}}$ is the heat
absorbed from the high temperature source and $\Delta Q_{\mathrm{L}}=\int_{0}^{\tau}dQ_{\mathrm{L}}$
the heat releases to the low temperature heat source. When the heat
engine works after many cycles, we can evaluate the heat engine's
performance with its efficiency $\eta$ and power $P$, namely,

\begin{equation}
\eta=\frac{\int_{0}^{t}dW}{-\int_{0}^{t}dQ_{\mathrm{H}}},P=\frac{\int_{0}^{t}dW}{t}
\end{equation}
where $t$ is the engine's total working time. For the size of these
two heat sources approach infinite as well as the heat capacity, i.e.,
$C_{\mathrm{H,L}}\rightarrow\infty$, their temperature will remain
constant as the initial time. In such case, the two heat sources service
as two heat bath, and the maximum efficiency of the heat engine is
bounded by the well-konwn Carnot efficiency

\begin{equation}
\eta_{\mathrm{C}}=1-\frac{T_{\mathrm{L}}}{T_{\mathrm{H}}},
\end{equation}
which can be achieved when the working substance works under a reversible
Carnot cycle with vanishing output power. However, for the heat sources
with finite size, their temperature will change as the heat engine
works. Specifically, after providing heat to the working substance,
$T_{\mathrm{H}}$ of the high-temperature heat source decreases. And
on the other hand, the low-temperature source's temperature $T_{\mathrm{L}}$
increases since the working substance releases heat to it. Here we
have assumed $C_{\mathrm{H,L}}>0$ since most physical systems have
positive heat capacity, and the extreme special situation where the
heat sources have negative heat capacity i.e., $C_{\mathrm{H,L}}<0$
will be discussed in Sec. \ref{sec:Heat-source-with-nagetive}. As
the time goes by, the heat engine will finally stop outputting work
when the temperature of the two heat sources become the same, namely,
$T_{\mathrm{H}}\left(t_{\mathrm{f}}\right)=T_{\mathrm{L}}\left(t_{\mathrm{f}}\right)$,
where $t_{\mathrm{f}}$ is introduced as the stop time of the heat
engine. The corresponding efficiency

\begin{equation}
\eta\left(t_{\mathrm{f}}\right)=\frac{\int_{0}^{t_{\mathrm{f}}}dW}{-\int_{0}^{t_{\mathrm{f}}}dQ_{\mathrm{H}}}
\end{equation}
is called the efficiency at maximum work (EMW) \citep{leff1987thermal,izumida2014work}
and we denote it as $\eta\left(t_{\mathrm{f}}\right)\equiv\eta_{\mathrm{EMW}}$
thereafter. Generally, the temperature change of the source per cycle
is $\Delta T_{\alpha}=\Delta Q_{\alpha}/C_{\alpha}$ ($\mathrm{\alpha=H},\mathrm{L}$)
with $\Delta Q_{\alpha}\propto C_{\mathrm{S}}\left(T_{\mathrm{H}}-T_{\mathrm{L}}\right)$,
thus the stop time $t_{\mathrm{f}}$ depends on the ratio of the heat
capacity of the sources and working substance as $t_{\mathrm{f}}\propto\mathrm{min}\left\{ C_{\mathrm{H}},C_{\mathrm{L}}\right\} /C_{\mathrm{S}}\tau$.
When the heat source and the working substance are about the same
size $\left(C_{\mathrm{S}}\sim C_{\mathrm{H}}\right)$, the heat engine
can only work for a few cycles or even less than one cycle, and the
heat engine in such circumstance does not have practical use value.
In the following discussion, we focus on the situation that the heat
engine can work with limited but sufficient cycles , i.e., $t_{\mathrm{f}}/\tau\gg1$
with $C_{\mathrm{S}}/C_{\alpha}\ll1$.

With the work of the heat engine keeps going on, noticing the temperature
difference of the two source become smaller and smaller, thus we have
\citep{izumida2014work}

\begin{equation}
\eta_{\mathrm{EMW}}=1+\frac{\int_{0}^{t_{\mathrm{f}}}dQ_{\mathrm{L}}}{\int_{0}^{t_{\mathrm{f}}}dQ_{\mathrm{H}}}<1+\frac{\int_{0}^{t_{\mathrm{f}}}T_{\mathrm{L}}dS_{\mathrm{L}}}{\int_{0}^{t_{\mathrm{f}}}T_{\mathrm{H}}dS_{\mathrm{H}}}\leq\eta_{\mathrm{C}},\label{eq:eta}
\end{equation}
where the equal sign on the right side only hold in the reversible
limit with $\int_{0}^{t_{\mathrm{f}}}dS_{\mathrm{L}}=-\int_{0}^{t_{\mathrm{f}}}S_{\mathrm{H}}$.
The above discussion implies that it is the finite heat capacity of
the heat sources limit the EMP the heat engine can achieve. Therefore,
the following two questions naturally raises: (i) What's the heat
engine's maximum EMP when working at two finite-size heat bath. (ii)
How the specific feature of the heat source's heat capacity affects
such EMP.We first rewrite the efficiency of Eq. (\ref{eq:eta}) in
term of $C_{\mathrm{H}}$ and $C_{\mathrm{L}}$ as

\begin{equation}
\eta_{\mathrm{EMW}}=1+\frac{\int_{0}^{t_{\mathrm{f}}}C_{\mathrm{L}}dT_{\text{\ensuremath{\mathrm{L}}}}}{\int_{0}^{t_{\mathrm{f}}}C_{\mathrm{H}}dT_{\text{\ensuremath{\mathrm{H}}}}},\label{eq:eta=00003DCT}
\end{equation}
where $\int_{0}^{t_{\mathrm{f}}}dQ_{\mathrm{\alpha}}=\int_{0}^{t_{\mathrm{f}}}dU_{\mathrm{\alpha}}=\int_{0}^{t_{\mathrm{f}}}C_{\mathrm{\alpha}}dT_{\text{\ensuremath{\mathrm{\alpha}}}}$
($\mathrm{\alpha=H},\mathrm{L}$) have been used for the two heat
sources. For a given physical system, heat capacity is generally the
function of temperature, i.e., $C_{\mathrm{H,L}}=C_{\mathrm{H,L}}\left(T\right)$
, and one can complete the integral in Eq. (\ref{eq:eta=00003DCT})
explicitly with the specific form of $C_{\mathrm{H,L}}\left(T\right)$.
Assuming the heat capacity of the sources follow the Debye's Law \citep{kittel1976introduction},
for example most of the crystal, thus $C\left(T\right)=$const in
the high temperature regime of $T/\Theta\gg1$, and $C\left(T\right)\propto T^{n}$
in the low-temperature regime of $T/\Theta\gg1$. Here $\Theta$ and
$n$ are the Debye temperature and the dimension of material respectively.

\subsection{\label{subsec:High-temperature-limit}High temperature regime}

\begin{figure}
\centering{}\includegraphics[width=8.5cm]{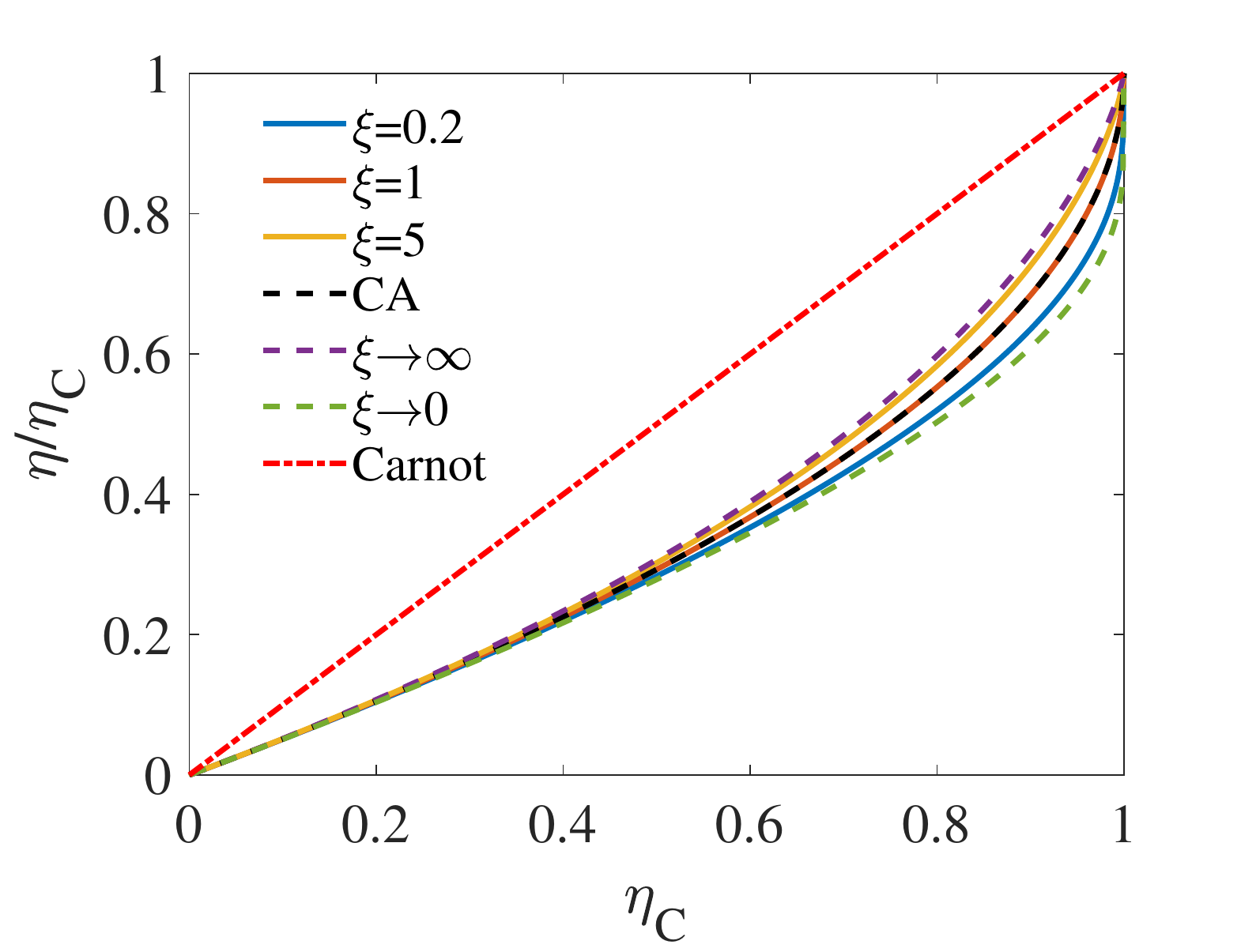}\caption{\label{fig:Maximum-efficiency-high-T-constantC}EMW $\eta_{\mathrm{EMW}}$
as the function of $\eta_{\mathrm{C}}$ with different $\xi=C_{\mathrm{L}}/C_{\mathrm{H}}.$
The curve relates to infinite low-temperature source size ( $\xi\rightarrow\infty$)
and infinite high-temperature source size ($\xi\rightarrow0$) are
given by Eq. (\ref{eq:eta_high_limit}). The other are plot with Eq.
(\ref{eq:eta_high}) .}
\end{figure}

In this case, the heat capacity only determined by the size (particle
number) of the source, such that Eq. (\ref{eq:eta=00003DCT}) is simplified
as 

\begin{equation}
\eta_{\mathrm{EMW}}=1-\frac{C_{\mathrm{L}}\left[T_{\text{\ensuremath{\mathrm{L}}}}\left(t_{\mathrm{f}}\right)-T_{\text{\ensuremath{\mathrm{L}}}}\right]}{C_{\mathrm{H}}\left[T_{\text{\ensuremath{\mathrm{H}}}}-T_{\text{\ensuremath{\mathrm{H}}}}\left(t_{\mathrm{f}}\right)\right]},\label{eq:eta-general}
\end{equation}
where$T_{\text{\ensuremath{\mathrm{L}}}}\left(t_{\mathrm{f}}\right)=T_{\text{\ensuremath{\mathrm{H}}}}\left(t_{\mathrm{f}}\right)\equiv T_{\mathrm{E}}$is
the equilibrium temperature of the two sources. The EMW is achieved
with the reversible cycle, in which no irreversible entropy is generated,
i.e., $\sum_{\mathrm{\alpha=H},\mathrm{L},\mathrm{S}}\int_{0}^{t_{\mathrm{f}}}S_{\alpha}=0$.
Noticing $C_{\mathrm{S}}/C_{\alpha}\ll1$, the entropy change of the
working substance $\int_{0}^{t_{\mathrm{f}}}S_{\mathrm{S}}$can be
ignored, then we have

\begin{equation}
\int_{0}^{t_{\mathrm{f}}}\left(\frac{dQ_{\text{\ensuremath{\mathrm{H}}}}}{T_{\text{\ensuremath{\mathrm{H}}}}}+\frac{dQ_{\text{\ensuremath{\mathrm{C}}}}}{T_{\text{\ensuremath{\mathrm{C}}}}}\right)=0,
\end{equation}
which is further written as

\begin{equation}
\int_{0}^{\tau}\frac{C_{\mathrm{L}}dT_{\text{\ensuremath{\mathrm{L}}}}}{T_{\text{\ensuremath{\mathrm{L}}}}}+\int_{0}^{\tau}\frac{C_{\mathrm{H}}dT_{\text{\ensuremath{\mathrm{H}}}}}{T_{\text{\ensuremath{\mathrm{H}}}}}=0\label{eq:reversible-condition}
\end{equation}
This is the reversible condition of the whole process. Finish the
integral in Eq. (\ref{eq:reversible-condition}), we find the equilibrium
temperature

\begin{equation}
T_{\mathrm{E}}=T_{\text{\ensuremath{\mathrm{H}}}}\left(1-\eta_{\mathrm{C}}\right)^{\frac{\xi}{\xi+1}},\label{eq:Te}
\end{equation}
where $\xi\equiv C_{\mathrm{L}}/C_{\mathrm{H}}$ is defined as the
heat capacity ratio between the low and high temperature heat source.
$\xi$ characterizes the asymmetry of two heat sources. By substituting
Eq. (\ref{eq:Te}) into Eq. (\ref{eq:eta-general}), the EMW is obtained
in terms of $\xi$ and $\eta_{\mathrm{C}}$ as

\begin{equation}
\eta_{\mathrm{EMW}}=1-\frac{\xi\left[\left(1-\eta_{\mathrm{C}}\right)^{\frac{\xi}{\xi+1}}-\left(1-\eta_{\mathrm{C}}\right)\right]}{1-\left(1-\eta_{\mathrm{C}}\right)^{\frac{\xi}{\xi+1}}}.\label{eq:eta_high}
\end{equation}
Here $\eta_{\mathrm{C}}=1-T_{\text{\ensuremath{\mathrm{L}}}}/T_{\text{\ensuremath{\mathrm{H}}}}$
is the Carnot efficiency determined by the initial temperature of
the sources. We illustrated $\eta_{\mathrm{max}}$ as the function
of $\eta_{\mathrm{C}}$ in Fig. \ref{fig:Maximum-efficiency-high-T-constantC}.
In three different limit case, namely, the symmetry case of $\xi=1$,
infinite low-temperature source size of $\xi\rightarrow\infty$, and
infinite high-temperature source size of $\xi\rightarrow0$, we simplify
Eq. (\ref{eq:eta_high}) as follows

\begin{equation}
\eta_{\mathrm{EMW}}=\begin{cases}
1-\sqrt{1-\eta_{\mathrm{C}}} & \xi=1\\
1+\left(\eta_{\mathrm{C}}^{-1}-1\right)\ln\left(1-\eta_{\mathrm{C}}\right) & \xi\rightarrow\infty\\
1+\eta_{\mathrm{C}}\ln^{-1}\left(1-\eta_{\mathrm{C}}\right) & \xi\rightarrow0
\end{cases}\label{eq:eta_high_limit}
\end{equation}
It is worth mentioning that, in the symmetry case, $\eta_{\mathrm{EMW}}$
is just the C-A efficiency\citep{CA}, which is the EMP of a symmetry
Carnot engine and has been obtained in many finite-time thermodynamics
model \citep{schmiedl2008efficiency,EspositoPRL2010,ZCTuCPB,johal2019many}.
The maximum EMW in this case is achieved with infinite low-temperature
source size of $\xi\rightarrow\infty$. The results of Eq. (\ref{eq:eta_high_limit})
have also been reported in \citep{ondrechen1983generalized} with
constant heat capacity. And we find these $\eta_{\mathrm{EMW}}$ in
different limits of $\xi$ have the same coefficient in the first
order of $\eta_{\mathrm{C}}$, namely, $\eta_{\mathrm{EMW}}=\eta_{\mathrm{C}}/2+O\left(\eta_{\mathrm{C}}^{2}\right)$,
which shares the same universality with EMP \citep{EspositoPRL2010,ZCTuCPB}

\subsection{\label{subsec:low-temperature-limit}low temperature regime}

\begin{figure}
\centering{}\includegraphics[width=8.5cm]{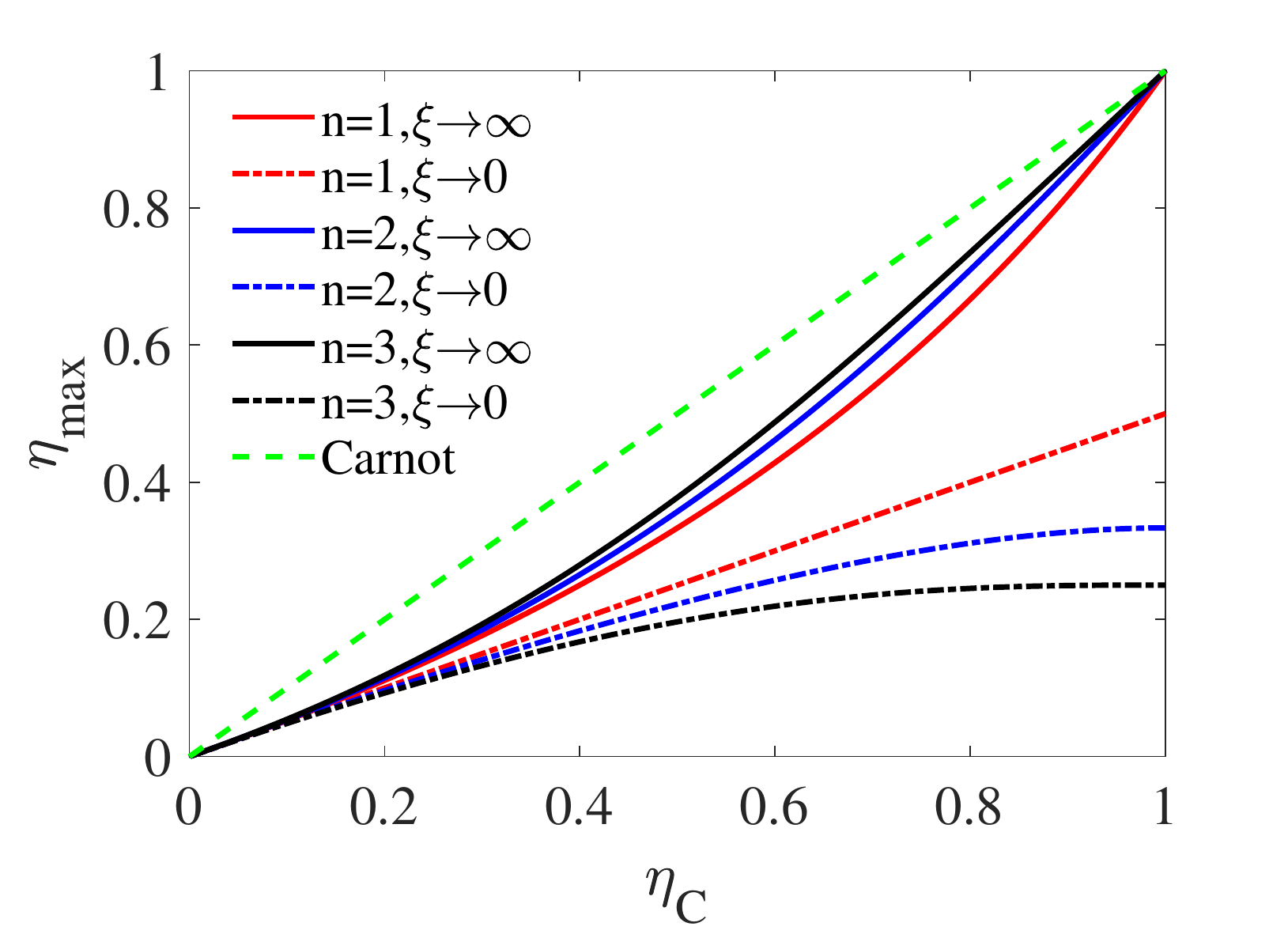}\caption{\label{fig. eff-n}EMW $\eta_{\mathrm{EMW}}$ as the function of $\eta_{\mathrm{C}}$
with different source dimension $n$. The curve relates to infinite
low-temperature source size ( $\xi\rightarrow\infty$) and infinite
high-temperature source size ($\xi\rightarrow0$) are given by Eq.
(\ref{eq:eta_max_lowT_xi_infinite}) and Eq. (\ref{eq:eta_max_lowT_xi0})
respectively. }
\end{figure}

Different from the above case within the high-temperature regime,
the capacity of the heat sources in the low-temperature regime of
$T/\Theta\ll1$, according to the Debye's law, follows $C_{\mathrm{H},\mathrm{L}}\left(T\right)=\Lambda_{\mathrm{H},\mathrm{L}}T^{n}$.
The reversible condition of Eq. (\ref{eq:reversible-condition}) in
this case becomes

\begin{equation}
\int_{0}^{t_{\mathrm{f}}}\Lambda_{\mathrm{L}}T_{\text{\ensuremath{\mathrm{L}}}}^{n-1}dT_{\text{\ensuremath{\mathrm{L}}}}+\int_{0}^{t_{\mathrm{f}}}\Lambda_{\mathrm{H}}T_{\text{\ensuremath{\mathrm{H}}}}^{n-1}dT_{\text{\ensuremath{\mathrm{H}}}}=0,\label{eq:reversible-condition-1}
\end{equation}
namely,

\begin{equation}
\Lambda_{\mathrm{L}}\left(T_{\text{\ensuremath{\mathrm{E}}}}^{n}-T_{\text{\ensuremath{\mathrm{L}}}}^{n}\right)=\Lambda_{\mathrm{H}}\left(T_{\text{\ensuremath{\mathrm{H}}}}^{n}-T_{\text{\ensuremath{\mathrm{E}}}}^{n}\right),
\end{equation}
which gives the equilibrium temperature

\begin{equation}
T_{\text{\ensuremath{\mathrm{E}}}}=\left(\frac{T_{\text{\ensuremath{\mathrm{H}}}}^{n}+\xi T_{\text{\ensuremath{\mathrm{L}}}}^{n}}{1+\xi}\right)^{\frac{1}{n}}\label{eq:Te_CT}
\end{equation}
with the capacity ratio reduces to $\xi=\Lambda_{\mathrm{L}}/\Lambda_{\mathrm{H}}$.
Combining Eqs. (\ref{eq:Te_CT}) and (\ref{eq:eta=00003DCT}), the
EMW at low temperature regime reads

\begin{equation}
\eta_{\mathrm{EMW}}=1-\frac{\xi\chi-\xi\left(1-\eta_{\mathrm{C}}\right)^{n+1}}{1-\chi},\label{eq:eta-Max-Low-temperature}
\end{equation}
where

\begin{equation}
\chi=\left[\frac{1+\xi\left(1-\eta_{\mathrm{C}}\right)^{n}}{1+\xi}\right]^{\frac{n+1}{n}}
\end{equation}
In the limit of infinite low-temperature source size ($\xi\rightarrow\infty$),
keeping the first order of $\xi^{-1}$ in $\chi$, we obtain

\begin{equation}
\eta_{\mathrm{EMW}}\left(\xi\rightarrow\infty\right)=1-\frac{n+1}{n}\left[1-\frac{\eta_{\mathrm{C}}}{1-\left(1-\eta_{\mathrm{C}}\right)^{n+1}}\right].\label{eq:eta_max_lowT_xi_infinite}
\end{equation}
On the other hand, in the case with infinite high-temperature source
size ($\xi\rightarrow0$), we expand $\chi$ up to the first order
of $\xi$ and find that

\begin{equation}
\eta_{\mathrm{EMW}}\left(\xi\rightarrow0\right)=1-\frac{n}{n+1}\left[\frac{1-\left(1-\eta_{\mathrm{C}}\right)^{n+1}}{1-\left(1-\eta_{\mathrm{C}}\right)^{n}}\right]\label{eq:eta_max_lowT_xi0}
\end{equation}
In Fig. \ref{fig. eff-n}, the EMW in these two limit cases are plotted
as the function of $\eta_{\mathrm{C}}$. The curves shows that $\eta_{\mathrm{EMW}}\left(\xi\rightarrow\infty\right)$,
the upper bound of $\eta_{\mathrm{EMW}}$, increases with the heat
source dimension $n$; while the lower bound of $\eta_{\mathrm{EMW}}$,
i.e., $\eta_{\mathrm{EMW}}\left(\xi\rightarrow0\right)$ decreases
with $n$. This means we can use high-dimension heat source with the
low-temperature source much lager than the high-temperature one to
realize higher efficiency.

Particularly, when the considered heat source is one-dimensional,
i.e., $n=1$, Eq. (\ref{eq:eta-Max-Low-temperature}) is directly
simplified as

\begin{equation}
\eta_{\mathrm{EMW}}\left(n=1\right)=\frac{\eta_{\mathrm{C}}}{2-\kappa\eta_{\mathrm{C}}},\label{eq:eta-max-low-n=00003D1}
\end{equation}
where $\kappa=\xi/\left(1+\xi\right),\kappa\in(0,\infty)$. Since
$\eta_{\mathrm{EMW}}\left(n=1\right)$ is a monotonically increasing
function of $\kappa$, we conclude that 

\begin{equation}
\eta_{\mathrm{U}}\equiv\frac{\eta_{\mathrm{C}}}{2}\leq\eta_{\mathrm{EMW}}\left(n=1\right)\leq\frac{\eta_{\mathrm{C}}}{2-\eta_{\mathrm{C}}}\equiv\eta_{\mathrm{L}}.\label{eq:eta+-}
\end{equation}
Interestingly, the upper and lower bound $\eta_{\mathrm{U,L}}$ here
are exactly the same as the bounds for the EMP obtained with different
finite-time heat engine models \citep{schmiedl2008efficiency,EspositoPRL2010,ZCTuCPB}.
Similar result has been found by \citep{johal2016near} with the energy-entropy
relation. In addition, in the limit of $\eta_{\mathrm{C}}\rightarrow1$,
the lower bound of $\eta_{\mathrm{EMW}}$ in Eq. (\ref{eq:eta_max_lowT_xi0})
is found to be only determined by the heat source dimension, i.e.,

\begin{equation}
\underset{\eta_{\mathrm{C}}\rightarrow1}{\lim}\eta_{\mathrm{EMW}}\left(\xi\rightarrow0\right)=\frac{1}{n+1}.\label{eq:eta_max_lowT_xi0-1}
\end{equation}
This phenomenon is observed in Fig. \ref{fig. eff-n}, where the intersection
of the three dash-dotted lines and $\eta_{\mathrm{C}}=1$ are, from
top to bottom, $\eta_{\mathrm{EMW}}=1/2$, $\eta_{\mathrm{EMW}}=1/3$,
and $\eta_{\mathrm{EMW}}=1/4$ respectively. Moreover, up to the first
order of $\eta_{\mathrm{C}}$, an universality is found the general
result of EMW both in the high and low temperature regime, namely,

\begin{equation}
\eta_{\mathrm{EMW}}=\frac{\eta_{\mathrm{C}}}{2}+O\left(\eta_{\mathrm{C}}^{2}\right).\label{eq:etaEMW-universality}
\end{equation}
The same universality has been discovered before for EMP\citep{EspositoPRL2010,ZCTuCPB}.

For those systems that do not satisfy the Debye's law, we can first
derive the heat capacity as the function of temperature from its energy
spectrum and the corresponding equilibrium population. Then taking
using of Eqs. (\ref{eq:Te_CT}) and (\ref{eq:reversible-condition})
to obtain the EMW in quasi-static limit. It should be mentioned that
the general result of Eq. (\ref{eq:eta-Max-Low-temperature}) is also
suitable for Fermi gas at low temperature and photon gas (black body
radiation) by taking $n=1$ and $n=3$ respectively. Since the capacity
of the former follows $C_{\mathrm{FG}}\left(T\right)\propto T$, while
$C_{\mathrm{PG}}\left(T\right)\propto T^{3}$ for the latter \citep{pathria1996statistical}.

\section{\label{sec:Efficiency-at-maximum}Finite-time performance of the
heat engine}

\begin{figure}
\begin{centering}
\includegraphics[width=8.5cm]{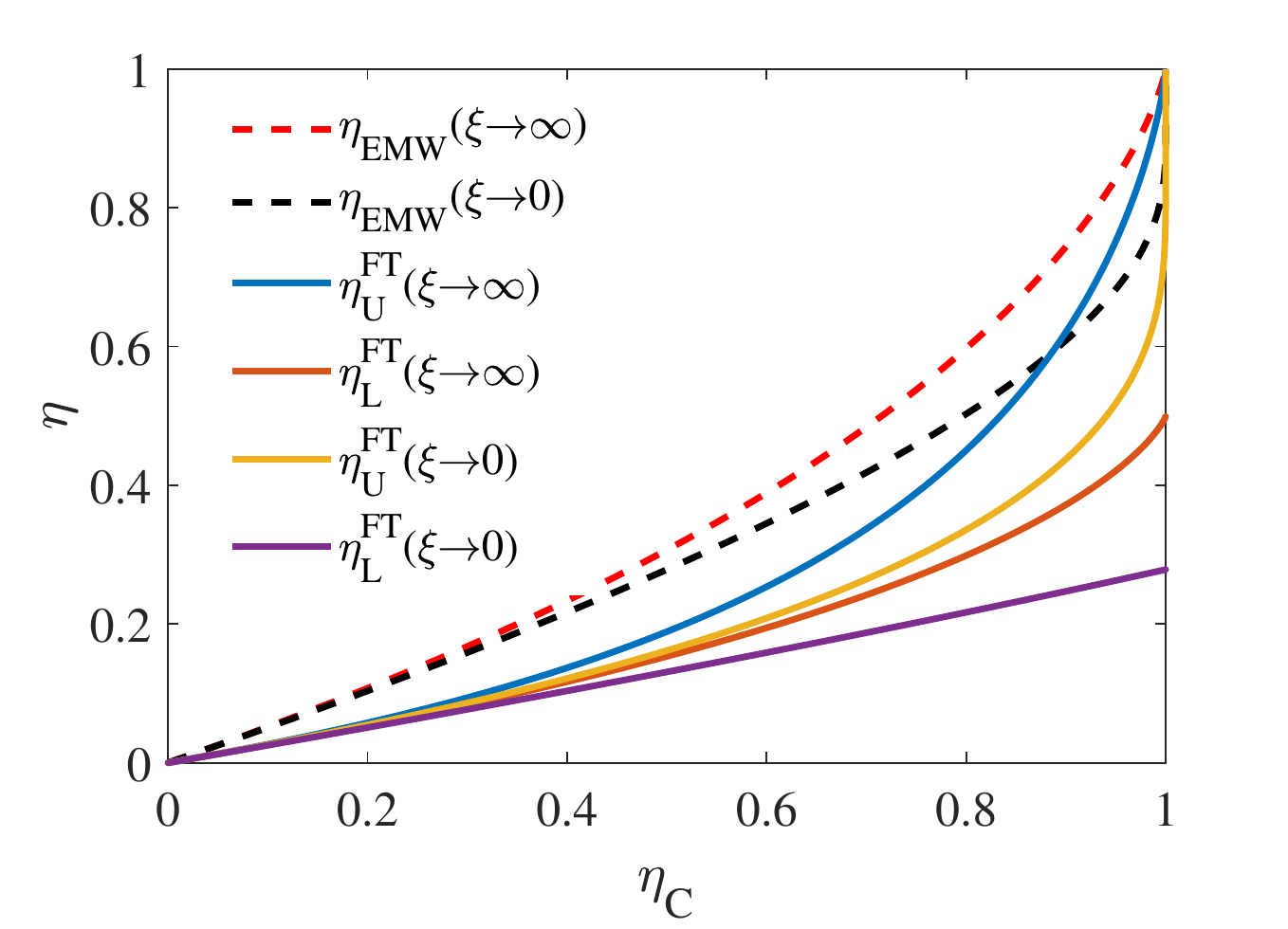}\caption{\label{fig:Efficiency-of-the-finite-time}Upper ($\eta_{\mathrm{U}}^{\mathrm{FT}}$)
and lower ($\eta_{\mathrm{L}}^{\mathrm{FT}}$) bounds for efficiency
of the heat engine in finite-time operation as the function of $\eta_{\mathrm{C}}$
in different limit of $\xi$ . $\eta_{\mathrm{U}}^{\mathrm{FT}}$
and $\eta_{\mathrm{L}}^{\mathrm{FT}}$ respectively corresponds to
$\gamma\rightarrow1$ and $\gamma\rightarrow0$. Here $\xi$ and $\gamma$
respectively characterize the asymmetry in size and dissipation of
the two heat sources. As the comparison, the red(black) dashed line
represent the upper (lower) bound of EMP in the reversible limit given
by Eq. (\ref{eq:eta_high_limit}) . Other are plotted with Eqs. (\ref{eq:xiinfr1}),
(\ref{eq:xiinfr0}), (\ref{eq:xi0r1}), and (\ref{eq:xi0r0}).}
\par\end{centering}
\end{figure}

In the previous section, we derive the EMW that heat engines can achieve
under the reversible condition, which is satisfied by a quasi-static
cycle with vanishing output power. In this section, we extend our
discussion in Sec. \ref{sec:heat-engine-working-finite source} to
the finite-time case\citep{izumida2014work,wang2014optimization}.
Unlike the optimization goal of Ref\citep{izumida2014work,wang2014optimization},
where the efficiency at maximum time-average power of the whole process
is studied, we focus on how efficient the engine can be when the power
of each cycle is maximized. We assume the engine can run $N$ cycle
from $t=0$ to the stop time $t=t_{\mathrm{f}}$, where $N\gg1$ but
finite as we mentioned before. The operation time of the $i$-th cycle
($i=1,2,...N$) begins at $t=t_{i}$ is denoted as $\tau_{i}=t_{i+1}-t_{i}$.
The efficiency of the whole process then reads

\begin{equation}
\eta=1-\frac{\sum_{i=1}^{N}\Delta Q_{\mathrm{C}}^{(i)}}{\sum_{i=1}^{N}\Delta Q_{\mathrm{H}}^{(i)}},\label{eq:eta-N-cycle}
\end{equation}
where $\Delta Q_{\mathrm{H}}^{(i)}=-\int_{t_{i}}^{t_{i+1}}dQ_{\mathrm{H}}$
and $\Delta Q_{\mathrm{C}}^{(i)}=\int_{t_{i}}^{t_{i+1}}dQ_{\mathrm{C}}$.
In the $i$-th cycle, the output power is $P^{(i)}=\left(\Delta Q_{\mathrm{H}}^{(i)}-\Delta Q_{\mathrm{C}}^{(i)}\right)\tau_{i}^{-1}$.
Below we try to obtain the upper and lower bounds for $\eta$ under
the condition that the maximum $P^{(i)}$ is achieved in each cycle.
Suppose the heat engine works in the Carnot-like cycle, which contains
two adiabatic processes and two finite-time isothermal process. In
the finite-time isothermal processes, the heat transfer generally
follows, in the the $i$-th cycle,

\begin{equation}
\Delta Q_{\mathrm{H}}^{(i)}=T_{\mathrm{H}}^{(i)}\left(\Delta S_{\mathrm{re}}^{(i)}-\Delta S_{\mathrm{\mathrm{irr},\mathrm{H}}}^{(i)}\right),
\end{equation}

\begin{equation}
\Delta Q_{\mathrm{C}}^{(i)}=T_{\mathrm{C}}^{(i)}\left(\Delta S_{\mathrm{re}}^{(i)}+\Delta S_{\mathrm{irr},\mathrm{C}}^{(i)}\right).
\end{equation}
Here, $T_{\mathrm{H}}^{(i)}$($T_{\mathrm{C}}^{(i)}$) is the temperature
of the high (low) temperature heat source, $\Delta S_{\mathrm{re}}^{(i)}$
is reversible entropy change, and $\Delta S_{\mathrm{\mathrm{irr},\mathrm{\alpha}}}^{(i)}$
$(\alpha=\mathrm{H,C})$ is the irreversible entropy generation relates
to the corresponding source. With the low-dissipation assumption,
as suggested first by \citep{EspositoPRL2010}, the irreversible entropy
follows the $1/\tau$ relation as $\Delta S_{\mathrm{\mathrm{irr},\mathrm{\alpha}}}^{(i)}=\Sigma_{\alpha}^{(i)}/\tau_{\alpha}^{(i)}$,
where $\Sigma_{\alpha}^{(i)}$ depends on the dissipative feature
of the working substance when contacting with the heat source \citep{Constraintrelationyhma}
and $\tau_{\alpha}^{(i)}$ is the operation time of the corresponding
process. Such relation has been proved for both classical \citep{MinEntropyProd}
and quantum system \citep{Constraintrelationyhma,yhmaoptimalcontrol}
and was recently observed in the experiment \citep{Ma2019IEGExp}.
Applying straightforward optimization of $P^{(i)}\left(\tau_{\mathrm{H}}^{(i)},\tau_{\mathrm{C}}^{(i)}\right)$
with respect to $\tau_{\mathrm{H}}^{(i)}$ and $\tau_{\mathrm{H}}^{(i)}$,
the EMP in the $i$-th cycle is obtained as \citep{EspositoPRL2010}(See
App. \ref{sec:Optimization-of-low-dissipation-APPA} for detailed
derivation)

\begin{equation}
\eta_{\mathrm{EMP}}^{(i)}=\frac{\eta_{C}^{(i)}}{2-\gamma^{\left(i\right)}\eta_{C}^{(i)}}
\end{equation}
with 

\begin{equation}
\gamma^{\left(i\right)}=\left(1+\sqrt{\frac{T_{\mathrm{C}}^{(i)}\Sigma_{\mathrm{C}}^{(i)}}{T_{\mathrm{H}}^{(i)}\Sigma_{\mathrm{H}}^{(i)}}}\right)^{-1}
\end{equation}
and $\eta_{C}^{(i)}=1-T_{\mathrm{C}}^{(i)}/T_{\mathrm{H}}^{(i)}$
being the Carnot efficiency determined by the temperature of the sources
in the $i$-th cycle. When the maximum power output for each cycle
is achieved, the efficiency of Eq. (\ref{eq:eta-N-cycle}),denoted
as $\eta^{\mathrm{FT}}$, becomes

\begin{equation}
\eta^{\mathrm{FT}}=\frac{\sum_{i=1}^{N}\eta_{\mathrm{EMP}}^{(i)}\Delta Q_{\mathrm{H}}^{(i)}}{\sum_{i=1}^{N}\Delta Q_{\mathrm{H}}^{(i)}},\label{eq:eta-N-cycle-EMP}
\end{equation}
which is a monotonically increasing function of $\eta_{\mathrm{EMP}}^{(i)}$.
Note that $\eta_{\mathrm{EMP}}^{(i)}$ is bounded by $\eta_{\mathrm{\pm}}^{(i)}$
as

\begin{equation}
\eta_{-}^{(i)}\equiv\frac{\eta_{\mathrm{C}}^{(i)}}{2}\leq\eta_{\mathrm{EMP}}^{(i)}\leq\frac{\eta_{\mathrm{C}}^{(i)}}{2-\eta_{\mathrm{C}}^{(i)}}\equiv\eta_{+}^{(i)},
\end{equation}
where the upper bound and lower bound are achieved respectively by
taking the limit of $\gamma\rightarrow1$ ($\Sigma_{\mathrm{C}}^{(i)}\rightarrow0$)
and $\gamma\rightarrow0$ ($\Sigma_{\mathrm{H}}^{(i)}\rightarrow0$).
Therefore,

\begin{equation}
\frac{\sum_{i=1}^{N}\eta_{-}^{(i)}\Delta Q_{\mathrm{H}}^{(i)}}{\sum_{i=1}^{N}\Delta Q_{\mathrm{H}}^{(i)}}\leq\eta^{\mathrm{FT}}\leq\frac{\sum_{i=1}^{N}\eta_{+}^{(i)}\Delta Q_{\mathrm{H}}^{(i)}}{\sum_{i=1}^{N}\Delta Q_{\mathrm{H}}^{(i)}}.\label{eq:up-low-eta}
\end{equation}
In in following, we will derive upper and lower bound for efficiency
of Eq. (\ref{eq:up-low-eta}). In the limit of $\gamma\rightarrow1$,
the upper bound of Eq. (\ref{eq:up-low-eta})

\begin{equation}
\frac{\sum_{i=1}^{N}\eta_{+}^{(i)}\Delta Q_{\mathrm{H}}^{(i)}}{\sum_{i=1}^{N}\Delta Q_{\mathrm{H}}^{(i)}}\equiv\eta_{\mathrm{U}}^{\mathrm{FT}}.\label{eq:eta-N-cycle-1}
\end{equation}
For simplicity, we only consider the heat source with constant heat
capacity in the following. As we studied In Sec. \ref{sec:heat-engine-working-finite source},
the EMW in the reversible situation is bounded as $\eta_{\mathrm{EMW}}\left(\xi\rightarrow0\right)\leq\eta_{\mathrm{EMW}}\leq\eta_{\mathrm{EMW}}\left(\xi\rightarrow\infty\right)$
(See Eq. (\ref{eq:eta_high_limit})). Therefore, to achieve higher
efficiency in finite time, we should focus on the case of $\xi\rightarrow\infty$,
where the low-temperature source is much lager than the high-temperature
one. Such that the cold source is kept at a constant temperature in
the whole the process, i.e., $T_{\mathrm{C}}^{(i)}=T_{\mathrm{C}}$
and the hot source temperature rises with time until $T_{\mathrm{H}}^{(N)}=T_{\mathrm{C}}$
and the heat engine stops working. Substituting Eq. (\ref{eq:dQ-eta})
into Eq. (\ref{eq:eta-N-cycle-1}), and replace the sum by integral
with $N\gg1$, we obtain

\begin{equation}
\eta_{\mathrm{U}}^{\mathrm{FT}}=\frac{\int_{0}^{t_{\mathrm{f}}}\frac{\eta_{\mathrm{C}}(t)}{2-\eta_{\mathrm{C}}(t)}dQ_{\mathrm{H}}}{\int_{0}^{t_{\mathrm{f}}}dQ_{\mathrm{H}}},\label{eq:eta-gamma1}
\end{equation}
which can by further simplified as, noticing $\eta_{\mathrm{C}}(t)=1-T_{\mathrm{C}}/T_{\mathrm{H}}\left(t\right)$
and $dQ_{\mathrm{H}}=C_{\mathrm{H}}dT_{\mathrm{H}}$,

\begin{equation}
\eta_{\mathrm{U}}^{\mathrm{FT}}=\frac{\int_{T_{\mathrm{H}}}^{T_{\mathrm{C}}}\frac{T_{\mathrm{H}}(t)-T_{\mathrm{C}}}{T_{\mathrm{H}}(t)+T_{\mathrm{C}}}dT_{\mathrm{H}}}{\int_{T_{\mathrm{H}}}^{T_{\mathrm{C}}}dT_{\mathrm{H}}},
\end{equation}
After finishing the integral, the upper bound for efficiency is finally
found as, in terms of the initial Carnot efficiency $\eta_{\mathrm{C}}$,

\begin{equation}
\eta_{\mathrm{U}}^{\mathrm{FT}}\left(\xi\rightarrow\infty\right)=1-\frac{2(1-\eta_{\mathrm{C}})}{\eta_{\mathrm{C}}}\ln\frac{2-\eta_{\mathrm{C}}}{2(1-\eta_{\mathrm{C}})}.
\end{equation}
Similarly, for $\gamma\rightarrow0$, by replacing $\eta_{+}\left(t\right)$
in Eq. (\ref{eq:eta-gamma1}) with $\eta_{-}\left(t\right)=\eta_{\mathrm{C}}(t)/2$,
we find 

\begin{equation}
\eta_{\mathrm{L}}^{\mathrm{FT}}\left(\xi\rightarrow\infty\right)=\frac{1}{2}\left[1+\left(\eta_{\mathrm{C}}^{-1}-1\right)\ln\left(1-\eta_{\mathrm{C}}\right)\right],
\end{equation}
which is exactly half of $\eta_{\mathrm{EMW}}\left(\xi\rightarrow\infty\right)$
in the reversible case. And the detailed derivations of other two
bounds in the limit of $\xi\rightarrow0$ are given in App. \ref{sec:Bounds-for-efficiency-APP}.
Here we make a brief summary of the efficiency bounds obtained in
different limit as follows

\begin{equation}
\eta_{\mathrm{U}}^{\mathrm{FT}}\left(\xi\rightarrow\infty\right)=1+\frac{2(1-\eta_{\mathrm{C}})}{\eta_{\mathrm{C}}}\ln\frac{2(1-\eta_{\mathrm{C}})}{2-\eta_{\mathrm{C}}},\label{eq:xiinfr1}
\end{equation}

\begin{equation}
\eta_{\mathrm{L}}^{\mathrm{FT}}\left(\xi\rightarrow\infty\right)=\frac{\eta_{\mathrm{EMW}}\left(\xi\rightarrow\infty\right)}{2},\label{eq:xiinfr0}
\end{equation}

\begin{equation}
\eta_{\mathrm{U}}^{\mathrm{FT}}\left(\xi\rightarrow0\right)=\frac{\eta_{\mathrm{EMW}}\left(\xi\rightarrow0\right)}{2-\eta_{\mathrm{EMW}}\left(\xi\rightarrow0\right)},\label{eq:xi0r1}
\end{equation}

\begin{equation}
\eta_{\mathrm{L}}^{\mathrm{FT}}\left(\xi\rightarrow0\right)=1+\frac{\eta_{\mathrm{C}}}{2}\ln^{-1}\left(1-\frac{\eta_{\mathrm{C}}}{2}\right),\label{eq:xi0r0}
\end{equation}
where $\eta_{\mathrm{EMP}}\left(\xi\rightarrow\infty\right)$ and
$\eta_{\mathrm{EMP}}\left(\xi\rightarrow0\right)$ are the corresponding
efficiency in the reversible limit as given by Eq. (\ref{eq:eta_high_limit}).
Interestingly, the dependence of $\eta$ on $\eta_{\mathrm{EMW}}$
in the limit of $\left(\xi\rightarrow\infty,\gamma\rightarrow0\right)$
and $\left(\xi\rightarrow0,\gamma\rightarrow1\right)$ follows the
same form as its corresponding counterpart in the infinite heat source
case. These efficiency in different limit of $\xi$ and $\gamma$
are illustrated in Fig. \ref{fig:Efficiency-of-the-finite-time} with
the solid lines, where the dashed lines are the upper and lower bounds
of EMP, i.e., $\eta_{\mathrm{EMP}}\left(\xi\rightarrow\infty\right)$
and $\eta_{\mathrm{EMP}}\left(\xi\rightarrow0\right)$. It's easily
to check that these efficiency follow the universality as

\begin{equation}
\eta=\frac{\eta_{\mathrm{C}}}{4}+O\left(\eta_{\mathrm{C}}^{2}\right).\label{eq:eta-uni}
\end{equation}
Comparing the above universality with Eq. (\ref{eq:etaEMW-universality}),
we can also write the universality of $\eta$ in term of $\eta_{\mathrm{EMW}}$as

\begin{equation}
\eta=\frac{\eta_{\mathrm{EMW}}}{2}+O\left(\eta_{\mathrm{EMW}}^{2}\right),\label{eq:eta-uni-1}
\end{equation}
which means that, up to the first order of $\eta_{\mathrm{C}}$, the
efficiency when the power of each cycle is maximized is just half
of the EMW. Such universality is also found for the efficiency at
maximum time-average power with linear irreversible heat engine under
the tight coupling condition \citep{izumida2014work,wang2014optimization}.

\section{\label{sec:Heat-source-with-nagetive}Black holes served as heat
sources}

In the previous section, We have discussed the situation that heat
engine operates to stop between two finite-size heat sources when
the high-temperature and low-temperature source reach the same temperature.
This actually based on the assumption that the heat capacity of the
heat sources are positive as we mentioned before, such that the temperature
of the high-temperature heat source is lowered and the temperature
of the low-temperature heat source is increased as the engine's working.
Although most physical systems have positive heat capacity, there
are indeed some systems with negative heat capacity, i.e., $C=\partial U/\partial T<0$,
such as black holes \citep{taylor1975introduction,bekenstein1980black}.
All the thermodynamic properties of a black hole only rely on its
mass $M$, angular momentum $J$, and charge $Q$, known as the three
hairs of black hole \citep{bekenstein1980black}. For simplicity,
we consider the Schwarzschild black hole, which only has one hair,
the mass. The internal energy and temperature of a Schwarzschild black
hole B with mass $M$ are respectively $U=M$ and $T=1/\left(8\pi M\right)$.
Here and after, we use the natural unit system. Therefore, the heat
capacity of B is $C=\partial U/\partial T=-8\pi M^{2}$. Obviously,
such heat capacity is negative and increases quadratically with the
black hole's mass.

\begin{figure}
\centering{}\includegraphics[width=8.5cm]{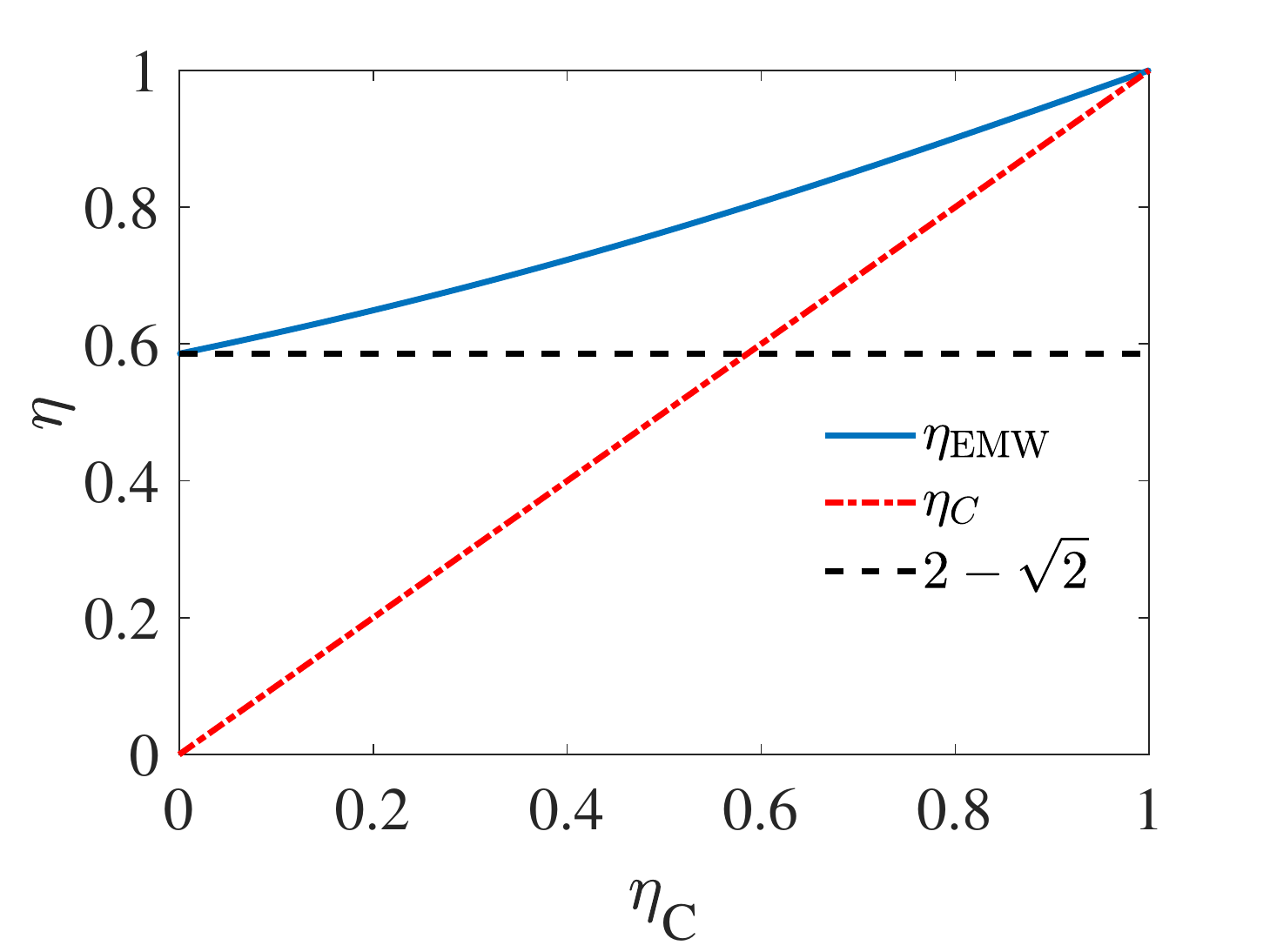}\caption{\label{fig:Maximum-efficiency-of-BH-Heat}EMW of a heat engine working
between two Schwarzschild black holes}
\end{figure}

Now we consider two Schwarzschild black holes of mass $M_{\text{\ensuremath{\mathrm{H}}}}$
and $M_{\text{\ensuremath{\mathrm{L}}}}$ served as high and low temperature
heat source respectively. Note the high temperature black hole has
smaller mass, than the low temperature one, namely, $M_{\text{\ensuremath{\mathrm{H}}}}<M_{\text{\ensuremath{\mathrm{L}}}}$.
The working substance reciprocates between the two black holes and
exchanges heat as well as output work, and we ignore the influence
of gravity on the cycle process in the following discussion. After
the heat engine absorbs heat from the high temperature black hole,
the mass of the high temperature black hole decreases, i.e., $M_{\text{\ensuremath{\mathrm{H}}}}\downarrow$
and then its temperature rises, namely $T_{\text{\ensuremath{\mathrm{H}}}}\uparrow$
; and when the heat is released to the low-temperature black hole,
the mass$M_{\text{\ensuremath{\mathrm{C}}}}\uparrow$ and the temperature
$T_{\text{\ensuremath{\mathrm{H}}}}\downarrow$ consequently. This
is exactly the opposite of what we discussed for the positive heat
capacity bath. Thus, the condition that the heat engine stops working
is no longer the temperature convergence of the two heat sources ($T_{\text{\ensuremath{\mathrm{L}}}}\left(t_{\mathrm{f}}\right)=T_{\text{\ensuremath{\mathrm{H}}}}\left(t_{\mathrm{f}}\right)\equiv T_{\mathrm{E}}$),
but the high temperature heat source, i.e., the smaller black hole,
is exhausted, namely, $M_{\text{\ensuremath{\mathrm{H}}}}\left(t_{\mathrm{f}}\right)=0$.
As a result, the efficiency of the heat engine work between these
two black holes follows

\begin{equation}
\eta=1-\frac{M_{\text{\ensuremath{\mathrm{L}}}}\left(t_{\mathrm{f}}\right)-M_{\text{\ensuremath{\mathrm{L}}}}}{M_{\text{\ensuremath{\mathrm{H}}}}}.\label{eq:eta_BH}
\end{equation}
Here we have assumed that the black holes only exchange heat with
the working substance without external energy transfer channels.We
still consider the reversible cycle for convenient, and the the reversible
condition of Eq. (\ref{eq:reversible-condition}) now becomes

\begin{equation}
\int_{0}^{t_{\mathrm{f}}}\frac{dM_{\text{\ensuremath{\mathrm{H}}}}}{1/\left(8\pi M_{\text{\ensuremath{\mathrm{H}}}}\right)}+\int_{0}^{t_{\mathrm{f}}}\frac{dM_{\text{\ensuremath{\mathrm{L}}}}}{1/\left(8\pi M_{\text{\ensuremath{\mathrm{L}}}}\right)}=0
\end{equation}
Then we obtain

\begin{equation}
M_{\text{\ensuremath{\mathrm{H}}}}^{2}+M_{\text{\ensuremath{\mathrm{L}}}}^{2}=M_{\text{\ensuremath{\mathrm{L}}}}^{2}\left(t_{\mathrm{f}}\right).\label{eq:MHMC}
\end{equation}
The above formula can also be derived with the conservation of black
hole area entropy with no information loss \citep{parikh2000hawking,zhang2009hidden,ma2018non,ma2018dark}
as

\begin{equation}
\sum_{\alpha=\mathrm{H,C}}S_{\mathrm{BH}}\left(M_{\text{\ensuremath{\mathrm{\alpha}}}}\right)=\sum_{\alpha=\mathrm{H,C}}S_{\mathrm{BH}}\left[M_{\text{\ensuremath{\mathrm{\alpha}}}}\left(t_{\mathrm{f}}\right)\right]
\end{equation}
with $S_{\mathrm{BH}}\left(M\right)=4\pi M^{2}$ being the Beckenstein-Hawking
Entropy. Combining Eqs. (\ref{eq:eta_BH}) and (\ref{eq:MHMC}), the
EMW reads

\begin{equation}
\eta_{\mathrm{EMW}}=\frac{\eta_{\mathrm{C}}+\sqrt{\left(1-\eta_{\mathrm{C}}\right)^{2}+1}}{1+\sqrt{\left(1-\eta_{\mathrm{C}}\right)^{2}+1}}\geq\eta_{\mathrm{C}},\label{eq:eta_max-BH}
\end{equation}
where

\begin{equation}
\eta_{\mathrm{C}}=1-\frac{T_{\text{\ensuremath{\mathrm{L}}}}}{T_{\text{\ensuremath{\mathrm{H}}}}}=1-\frac{M_{\text{\ensuremath{\mathrm{H}}}}}{M_{\text{\ensuremath{\mathrm{C}}}}}
\end{equation}
is the Carnot efficiency defined by the initial mass of the two black
holes. As demonstrated by Eq. (\ref{eq:eta_BH}), the maximum $\eta_{\mathrm{max}}$
can surpass the initial Carnot efficiency due to the unusual nature
of negative heat capacity of black hole. $\eta_{\mathrm{EMW}}$ as
the function of $\eta_{\mathrm{C}}$ is plotted in Fig. \ref{fig:Maximum-efficiency-of-BH-Heat}.
We should emphasize here that this result does not violate the second
law of thermodynamics, because the temperature of the two black holes
are not constant, but has an increasing temperature difference as
the heat engine works. Moreover, compared with the result relates
to positive heat capacity sources in Eqs. (\ref{eq:eta_high}) and
(\ref{eq:eta-Max-Low-temperature}), the EMP of Eq. (\ref{eq:eta_max-BH})
only rely on the initial Carnot efficiency without relying on other
parameters. We can regard this phenomenon as the embodiment of the
No-hair theorem \citep{bekenstein1980black}in efficiency for heat
engines operating between black holes.

When the initial mass of the two black holes are the same, i.e., $T_{\text{\ensuremath{\mathrm{L}}}}=T_{\text{\ensuremath{\mathrm{H}}}}$and
$\eta_{\mathrm{C}}=0$, the heat engine can be driven with some energy
fluctuation between the working substance and one of the black hole
due to the Hawing radiation process \citep{hawking1974black,hawking1975particle},
in such circumstance, one has

\begin{equation}
\underset{\eta_{\mathrm{C}}\rightarrow0}{\lim}\eta_{\mathrm{EMW}}=2-\sqrt{2}.
\end{equation}
This is two times of the maximum energy release rate $\mu=\left(2-\sqrt{2}\right)/2\approx0.29$
of two black holes' emerging process which was first derived by Hawking\citep{hawking1971gravitational}.
We note that there are studies that have linked black holes to heat
engines\citep{johnson2014holographic,hendi2018black,wei2019charged}
. However, to our best knowledge, these works mainly consider the
black holes as the working substance rather than heat sources, and
do not consider the finite-size effect.

Besides the black hole, the negative heat capacity also been observed
in some Cluster of atom system with phase transition \citep{d2000negative,schmidt2001negative,reyes2003negative}.
The discussion in this section can be extends to these systems, we
hope to use these novel materials to achieve high energy conversion
efficiency in the near future.

\section{\label{sec:Conclusion}Conclusion and discussion}

In summary, we studied the efficiency of a heat engine working between
two finite-size heat source in both quasi-static and finite-time cases.
The effect of the heat capacity of the finite-size heat source on
the engine's efficiency is clarified. When the engine operates in
quasi-static cycles, with the assumption that the source's capacity
follows the Debye's law, we obtained the corresponding efficiency
at maximum work (EMW) in the high and low temperature regime, as given
by Eq. (\ref{eq:eta_high}) and Eq. (\ref{eq:eta-Max-Low-temperature})
respectively. In addition, we also find some bounds for such EMW in
different circumstances {[}Eq. (\ref{eq:eta_high_limit}) of high
temperature regime and Eqs. (\ref{eq:eta_max_lowT_xi_infinite}) and
(\ref{eq:eta_max_lowT_xi0}) for low temperature{]}. We proved that,
in the low temperature case, with the limit of $\eta_{\mathrm{C}}\rightarrow1$,
the lower bound of $\eta_{\mathrm{EMW}}$ only determined by the heat
source's dimension, i.e., $\underset{\eta_{\mathrm{C}}\rightarrow1}{\lim}\eta_{\mathrm{EMW}}\left(\xi\rightarrow0\right)=(n+1)^{-1}$.
And for the 1-dimensional sources, the upper and lower bound $\eta_{\mathrm{U}}=\eta_{\mathrm{C}}(2-\eta_{\mathrm{C}})^{-1}$
and $\eta_{\mathrm{L}}=\eta_{\mathrm{C}}/2$ of $\eta_{\mathrm{EMW}}$
are exactly the same as the bounds for the efficiency at maximum power
(EMP) obtained with several finite-time heat engine models \citep{schmiedl2008efficiency,EspositoPRL2010,ZCTuCPB}.

For the heat engine working within finite time, we modeled the engine
as a low-dissipation Carnot-like engine and studied the efficiency
with the output power of each cycle is maximized. A series of bounds
for the efficiency are obtained in Eqs. (\ref{eq:xiinfr1}), (\ref{eq:xiinfr0}),
(\ref{eq:xi0r1}), and (\ref{eq:xi0r0}) and plotted in Fig. \ref{fig:Efficiency-of-the-finite-time},
where the overall upper bound is achieved with $\xi\rightarrow\infty,\gamma\rightarrow1$,
namely, the size of cold source is much lager than that of the hot
one while the dissipation of the cold source approaches vanishing.
An universality is found for all these bounds obtained with finite-size
sources and finite-time as $\eta=\eta_{\mathrm{C}}/4+O\left(\eta_{\mathrm{C}}^{2}\right)$,
where the coefficient of $\eta_{\mathrm{C}}$'s first order is just
half of that of the EMP.

Although we have discussed the effect of asymmetry in the size of
the hot and cold sources on the efficiency, the results are obtained
for the heat capacity function rely on temperature of the sources
share the same form. Considering that the high and low temperature
heat sources have different temperature dependence, such as cold source's
heat capacity varies with temperature following the power law while
the capacity of the hot source heat capacity remains constant, etc.,
is a potential direction for the optimization of the heat engine's
performance. In addition, the effect of phase transition (PT) is also
worth exploring in future study, as the heat capacity of sources with
PT may have completely different characteristics in different phases
due to temperature changes.

In the last part of this paper, we studied an unusual case where the
sources have negative heat capacity. Using black hole as a demonstration,
we obtain the EMW for a heat engine working between two Schwarzschild
black holes as shown in Eq. (\ref{fig:Maximum-efficiency-of-BH-Heat}).
The EMP is found always higher than the initial Carnot efficiency
defined by the initial mass of the two black holes. But we emphasize
that this does not violate the second law of thermodynamics, since
the temperature of the two black hole are not constant but has an
increasing temperature difference as the heat engine operates due
to the negative heat capacity of the sources. In addition, even when
the initial mass of the two black holes are the same, the heat engine
can be driven with some energy fluctuation between the working substance
and one of the black hole from the Hawing radiation process. In this
situation, the EMW is proved to be $\eta_{\mathrm{EMW}}=2-\sqrt{2}=2\mu$,
where $\mu=\left(2-\sqrt{2}\right)/2\approx0.29$ is the maximum energy
release rate of two black holes' emerging process discovered by S.
Hawing \citep{hawking1971gravitational}. The discussions about this
issue in this paper are limited to quasi-static heat engines working
between the Schwarzschild black holes, and the black holes only exchange
heat with the working substance without external energy transfer.
Considering the finite-time effect, the non-negligible energy loss
due to the Hawking radiation, the cases with other types of black
holes, and the gravitational effect on the efficiency of the heat
engine will be a series of interesting and challenging tasks, which
will be investigated in our further studies. Moreover, since there
have been some experimental reports on negative heat capacity materials
\citep{d2000negative,schmidt2001negative,reyes2003negative}, it is
feasible to take use of these materials as heat sources providing
energy for the heat engines to test our predictions.
\begin{acknowledgments}
This work is supported by the NSFC (Grants No. 11534002 and No. 11875049),
the NSAF (Grant No. U1730449 and No. U1530401), and the National Basic
Research Program of China (Grants No. 2016YFA0301201 and No. 2014CB921403).
The author thanks Yun-He Zhao of Capital Normal University High School
for carefully proofreading of this paper.
\end{acknowledgments}

\bibliographystyle{apsrev}
\bibliography{finitesize}

\appendix

\section{\label{sec:Optimization-of-low-dissipation-APPA}Optimization of
low-dissipation Carnot-like engine }

In this Appendix, we briefly show the optimization of the low-dissipation
Carnot-like engine introduced in Sec. \ref{sec:Efficiency-at-maximum}
,we follow the same optimization method in Ref. \citep{EspositoPRL2010}.
For such engine working in the $i$-th Carnot-like cycle, the heat
transfer in the two finite-time isothermal processes read

\begin{equation}
\Delta Q_{\mathrm{H}}^{(i)}=T_{\mathrm{H}}^{(i)}\left(\Delta S_{\mathrm{re}}^{(i)}-\frac{\Sigma_{\mathrm{H}}^{(i)}}{\tau_{\mathrm{H}}^{(i)}}\right),\label{eq:QH}
\end{equation}
and

\begin{equation}
\Delta Q_{\mathrm{C}}^{(i)}=T_{\mathrm{C}}^{(i)}\left(\Delta S_{\mathrm{re}}^{(i)}+\frac{\Sigma_{\mathrm{C}}^{(i)}}{\tau_{\mathrm{C}}^{(i)}}\right),\label{eq:QC}
\end{equation}
where, $T_{\mathrm{H}}^{(i)}$($T_{\mathrm{C}}^{(i)}$) is the temperature
of the high (low) temperature heat source, $\Delta S_{\mathrm{re}}^{(i)}$
is reversible entropy change, $\Sigma_{\alpha}^{(i)}$ depends on
the dissipative nature of the working substance when contacting with
the heat source and $\tau_{\alpha}^{(i)}$ is the corresponding operation
time. Thus, the efficiency and power of the engine for the $i$-th
cycle follows

\begin{align}
\eta^{(i)} & =\frac{\Delta Q_{\mathrm{H}}^{(i)}-\Delta Q_{\mathrm{C}}^{(i)}}{\Delta Q_{\mathrm{H}}^{(i)}}\\
 & =\frac{\left(T_{\mathrm{H}}^{(i)}-T_{\mathrm{C}}^{(i)}\right)\Delta S_{\mathrm{re}}^{(i)}-\frac{\Sigma_{\mathrm{H}}^{(i)}}{\tau_{\mathrm{H}}^{(i)}}-\frac{\Sigma_{\mathrm{C}}^{(i)}}{\tau_{\mathrm{C}}^{(i)}}}{T_{\mathrm{H}}^{(i)}\left(\Delta S_{\mathrm{re}}^{(i)}-\frac{\Sigma_{\mathrm{H}}^{(i)}}{\tau_{\mathrm{H}}^{(i)}}\right)}\label{eq:etai}
\end{align}
and

\begin{align}
P^{(i)} & =\frac{\Delta Q_{\mathrm{H}}^{(i)}-\Delta Q_{\mathrm{C}}^{(i)}}{\tau_{\mathrm{H}}^{(i)}+\tau_{\mathrm{C}}^{(i)}}\\
 & =\frac{\left(T_{\mathrm{H}}^{(i)}-T_{\mathrm{C}}^{(i)}\right)\Delta S_{\mathrm{re}}^{(i)}-\frac{\Sigma_{\mathrm{H}}^{(i)}}{\tau_{\mathrm{H}}^{(i)}}-\frac{\Sigma_{\mathrm{C}}^{(i)}}{\tau_{\mathrm{C}}^{(i)}}}{\tau_{\mathrm{H}}^{(i)}+\tau_{\mathrm{C}}^{(i)}}
\end{align}
Here the operation time in the two adiabatic processes are ignored
\citep{EspositoPRL2010}. The maximum power of each cycle is obtained
by setting the derivatives of $P^{(i)}=P^{(i)}\left(\tau_{\mathrm{H}}^{(i)},\tau_{\mathrm{C}}^{(i)}\right)$
with respect to $\tau_{\mathrm{H}}^{(i)}$ and $\tau_{\mathrm{C}}^{(i)}$
equal to zero. Thus we find the corresponding times for the engine
working at maximum power as

\begin{equation}
\tau_{\mathrm{H}}^{(i)}=2\frac{T_{\mathrm{H}}^{(i)}\Sigma_{\mathrm{H}}^{(i)}}{\left(T_{\mathrm{H}}^{(i)}-T_{\mathrm{C}}^{(i)}\right)\Delta S_{\mathrm{re}}^{(i)}}\left(1+\sqrt{\frac{T_{\mathrm{C}}^{(i)}\Sigma_{\mathrm{C}}^{(i)}}{T_{\mathrm{H}}^{(i)}\Sigma_{\mathrm{H}}^{(i)}}}\right)\label{eq:tauH}
\end{equation}
and

\begin{equation}
\tau_{\mathrm{C}}^{(i)}=\tau_{\mathrm{H}}^{(i)}\sqrt{\frac{T_{\mathrm{C}}^{(i)}\Sigma_{\mathrm{C}}^{(i)}}{T_{\mathrm{H}}^{(i)}\Sigma_{\mathrm{H}}^{(i)}}}\label{eq:tauC}
\end{equation}
Substituting Eqs. (\ref{eq:tauH}) and (\ref{eq:tauC}) into Eq. (\ref{eq:etai}),
the efficiency at maximum power is obtained as
\begin{equation}
\eta_{\mathrm{EMP}}^{(i)}=\frac{\eta_{C}^{(i)}}{2-\gamma^{\left(i\right)}\eta_{C}^{(i)}},
\end{equation}
where

\begin{equation}
\gamma^{\left(i\right)}=\left(1+\sqrt{\frac{T_{\mathrm{C}}^{(i)}\Sigma_{\mathrm{C}}^{(i)}}{T_{\mathrm{H}}^{(i)}\Sigma_{\mathrm{H}}^{(i)}}}\right)^{-1}
\end{equation}
and $\eta_{C}^{(i)}=1-T_{\mathrm{C}}^{(i)}/T_{\mathrm{H}}^{(i)}$
is the Carnot efficiency determined by the temperature of the sources
in the $i$-th cycle

\section{\label{sec:Bounds-for-efficiency-APP}Bounds for efficiency in finite
time in the limit of $\xi\rightarrow0$}

In this Appendix, we derive the upper and lower bounds for efficiency
of heat engine working between finite-size source within finite time
in the limit of $\xi\rightarrow0$. In this case, the cold source
is much smaller than the hot one, thus $T_{\mathrm{H}}^{(i)}=T_{\mathrm{H}}$
and $T_{\mathrm{C}}^{(N)}=T_{\mathrm{H}}$ at $t=t_{\mathrm{f}}$.
In the limit of $\gamma\rightarrow1$, the efficiency of Eq. (\ref{eq:eta-N-cycle-EMP})
becomes 

\begin{equation}
\eta=\frac{\sum_{i=1}^{N}\frac{\eta_{\mathrm{C}}^{(i)}}{2-\eta_{\mathrm{C}}^{(i)}}\Delta Q_{\mathrm{H}}^{(i)}}{\sum_{i=1}^{N}\Delta Q_{\mathrm{H}}^{(i)}}.\label{eq:eta_r=00003D1_xiinf}
\end{equation}
Note that in this limit, Eqs. (\ref{eq:tauH}) and (\ref{eq:tauC})
respectively reduce to $\tau_{\mathrm{H}}^{(i)}=2\Sigma_{\mathrm{H}}^{(i)}/(\eta_{\mathrm{C}}^{(i)}\Delta S_{\mathrm{re}}^{(i)})$
and $\tau_{\mathrm{C}}^{(i)}=0$, substituting which into Eq. (\ref{eq:QH})
and Eq. (\ref{eq:QC}), we find the relation between heat transfer
and reversible entropy change as

\begin{equation}
\Delta Q_{\mathrm{H}}^{(i)}=T_{\mathrm{H}}^{(i)}\Delta S_{\mathrm{re}}^{(i)}\left(1-\frac{\eta_{\mathrm{C}}^{(i)}}{2}\right).\label{eq:dQ-eta}
\end{equation}
and

\begin{equation}
\Delta Q_{\mathrm{C}}^{(i)}=T_{\mathrm{C}}^{(i)}\Delta S_{\mathrm{re}}^{(i)}.\label{eq:dQ-eta-1}
\end{equation}
Using Eqs. (\ref{eq:dQ-eta}) and (\ref{eq:dQ-eta-1}), and replace
the sum by integral with $N\gg1$, Eq. (\ref{eq:eta_r=00003D1_xiinf})
is simplified as

\begin{align}
\eta & =\frac{\int_{0}^{t_{\mathrm{f}}}\frac{\eta_{\mathrm{C}}(t)}{2-\eta_{\mathrm{C}}(t)}\left[1-\frac{\eta_{\mathrm{C}}(t)}{2}\right]T_{\mathrm{H}}dS_{\mathrm{re}}}{\int_{0}^{t_{\mathrm{f}}}\left[1-\frac{\eta_{\mathrm{C}}\left(t\right)}{2}\right]T_{\mathrm{H}}dS_{\mathrm{re}}}\label{eq:eta-gamma1-1}\\
 & =\frac{\int_{0}^{t_{\mathrm{f}}}\eta_{\mathrm{C}}(t)dS_{\mathrm{re}}}{\int_{0}^{t_{\mathrm{f}}}\left[2-\eta_{\mathrm{C}}(t)\right]dS_{\mathrm{re}}}\\
 & =\frac{\int_{T_{\mathrm{L}}}^{T_{\mathrm{H}}}\eta_{\mathrm{C}}(t)\left(\frac{C_{\mathrm{L}}dT_{\mathrm{L}}}{T_{\mathrm{L}}}\right)}{\int_{T_{\mathrm{L}}}^{T_{\mathrm{H}}}\left[2-\eta_{\mathrm{C}}(t)\right]\left(\frac{C_{\mathrm{L}}dT_{\mathrm{L}}}{T_{L}}\right)}\\
 & =\frac{\int_{T_{\mathrm{L}}}^{T_{\mathrm{H}}}\left(\frac{T_{\mathrm{H}}}{T_{\mathrm{L}}}-1\right)dT_{\mathrm{L}}}{\int_{T_{\mathrm{L}}}^{T_{\mathrm{H}}}\left(\frac{T_{\mathrm{H}}}{T_{\mathrm{L}}}+1\right)dT_{\mathrm{L}}}\label{eq:eta-gamma=00003D1-xi0}
\end{align}
By straightforward calculation, we have

\begin{equation}
\eta_{\mathrm{U}}^{\mathrm{FT}}\left(\xi\rightarrow0\right)=1-\frac{2\eta_{\mathrm{C}}}{\eta_{\mathrm{C}}-\ln\left(1-\eta_{\mathrm{C}}\right)},
\end{equation}
which can be re-expressed by $\eta_{\mathrm{max}}\left(\xi\rightarrow0\right)$
as

\begin{equation}
\eta_{\mathrm{U}}^{\mathrm{FT}}\left(\xi\rightarrow0\right)=\frac{\eta_{\mathrm{max}}\left(\xi\rightarrow0\right)}{2-\eta_{\mathrm{max}}\left(\xi\rightarrow0\right)}
\end{equation}
This is the upper bound for $\xi\rightarrow0$ we illustrated in Eq.
(\ref{eq:xi0r1}). On the other hand, for $\gamma\rightarrow0$, Eq.
(\ref{eq:eta-gamma1}) becomes, by replacing $\eta_{+}\left(t\right)$
with $\eta_{-}\left(t\right)=\eta_{\mathrm{C}}(t)/2$

\begin{align}
\eta & =\frac{\int_{0}^{t_{\mathrm{f}}}\frac{\eta_{\mathrm{C}}(t)}{2}T_{\mathrm{H}}dS_{\mathrm{re}}}{\int_{0}^{t_{\mathrm{f}}}T_{\mathrm{H}}dS_{\mathrm{re}}}\label{eq:eta-gamma0-xi0}\\
 & =\frac{\int_{T_{\mathrm{L}}}^{T_{\mathrm{H}}}\frac{\eta_{\mathrm{C}}(t)}{2}T_{\mathrm{H}}\left[\frac{2-2\eta_{\mathrm{C}}(t)}{2-\eta_{\mathrm{C}}(t)}\right]\left(\frac{C_{\mathrm{L}}dT_{\mathrm{L}}}{T_{\mathrm{L}}}\right)}{\int_{T_{\mathrm{L}}}^{T_{\mathrm{H}}}T_{\mathrm{H}}\left[\frac{2-2\eta_{\mathrm{C}}(t)}{2-\eta_{\mathrm{C}}(t)}\right]\left(\frac{C_{\mathrm{L}}dT_{\mathrm{L}}}{T_{\mathrm{L}}}\right)}\\
 & =\frac{\int_{T_{\mathrm{L}}}^{T_{\mathrm{H}}}\frac{T_{\mathrm{H}}-T_{\mathrm{L}}}{T_{\mathrm{H}}+T_{\mathrm{L}}}dT_{\mathrm{L}}}{2\int_{T_{\mathrm{L}}}^{T_{\mathrm{H}}}\frac{1}{T_{\mathrm{H}}+T_{\mathrm{L}}}dT_{\mathrm{L}}},\label{eq:eta-lower-r=00003D0=00003Dxi0}
\end{align}
where we have used

\begin{equation}
\Delta Q_{\mathrm{H}}^{(i)}=T_{\mathrm{H}}^{(i)}\Delta S_{\mathrm{re}}^{(i)}\label{eq:dQ-eta-2}
\end{equation}
and

\begin{equation}
\Delta Q_{\mathrm{C}}^{(i)}=T_{\mathrm{C}}^{(i)}\Delta S_{\mathrm{re}}^{(i)}\left(\frac{2-\eta_{\mathrm{C}}^{(i)}}{2-2\eta_{\mathrm{C}}^{(i)}}\right),\label{eq:dQ-eta-1-1}
\end{equation}
in the limit of $\gamma\rightarrow0$. Finish the integral of Eq.
(\ref{eq:eta-lower-r=00003D0=00003Dxi0}), we obtain

\begin{equation}
\eta_{\mathrm{L}}^{\mathrm{FT}}\left(\xi\rightarrow0\right)=1+\frac{\eta_{\mathrm{C}}}{2}\ln^{-1}\left(1-\frac{\eta_{\mathrm{C}}}{2}\right),
\end{equation}
which is the lower bound $\eta^{\mathrm{FT}}$of in the limit of $\xi\rightarrow0$
as illustrated in Eq. (\ref{eq:xi0r1}).

\end{document}